\documentclass[preprint2,10pt,a4paper]{aastex}
\usepackage{amsmath}
\usepackage{graphicx}
\usepackage{multicol}
\begin{document}
\title{On the excitation of f-modes and torsional modes by magnetar giant flares}
\author{Yuri Levin$^{1,2}$ \& Maarten van Hoven$^{2,3}$}
\affil{$^1$School of Physics and CSPA, Monash University, VIC 3800 Australia, and}
\affil{$^2$Leiden University, Leiden Observatory, Niels Bohrweg 2, 2300 RA Leiden, the Netherlands}
\affil{$^3$Leiden University, Lorentz Institute, Niels Bohrweg 2, 2300 RA Leiden, the Netherlands}
\email{yuri.levin@monash.edu.au, vhoven@strw.leidenuniv.nl}
\begin{abstract} \noindent
Magnetar giant flares may excite vibrational modes of  neutron stars. Here
we compute an estimate of  initial post-flare amplitudes of both the torsional modes 
in the magnetar's crust and of the global f-modes. We show that while the torsional
crustal modes can be strongly excited, only a small fraction of the flare's energy
is converted directly into the lowest-order f-modes. For a
conventional model of a magnetar, with the external magnetic field of $\sim 10^{15}$G, 
the gravitational-wave detection of these f-modes 
 with  advanced LIGO is unlikely.

\end{abstract}
\section{Introduction}
The gamma- and x-ray  flares from Soft Gamma Repeaters (SGRs; Mazetz et al.~1979, Hurley et al.~1998, 2004) 
are believed 
to be powered by a sudden release of magnetic energy stored in their host magnetars (Thompson \& Duncan 1995). An SGR flare may excite
vibrational modes of a magnetar (Duncan 1998). Indeed, torsional oscillations of a magnetar
provide an attractive explanation  for some of the quasi-periodic oscillations (QPOs)
observed in the tails of giant flares (Barat et al.~1983, Israel et al.~2005, Strohmayer \& Watts 2005,
van Hoven \& Levin 2011, Gabler et al.~2011, Colaiuda \& Kokkotas 2011). Moreover, there is now
some evidence for QPOs during the normal, non-giant flares in SGR 1806-20 (El-Mezeini \& Ibrahim 2010). 

Excitation of low-order f-modes is also of considerable interest, because of the f-modes' strong
coupling to potentially detectable gravitational radiation. The 
sensitivity of the ground-based gravitational-wave interferometers has dramatically improved
over the last 5 years (Abott et al.~2009a, Acernese et al.~2008), and interesting upper limits on the f-mode gravitational- 
wave emission from the 2004 SGR 1806-20 giant flare, a possible 2009  
SGR 1550--5418 giant flare, and several less energetic bursts
 have recently been obtained (Abott et al.~2008, Abott et al.~2009b, Abadie et al.~2010, see also Kalmus et al.~2009).
Advanced LIGO and VIRGO are expected to become operational in the next 5-7 years, and it is of interest
to predict the strength of  expected gravitational-wave signal from  future giant flares.
In this paper, we compute a theoretical estimate for the  amplitude of the
torsional and f-modes expected to be excited in a giant flare. We show that only a small fraction
of the flare energy is expected to be pumped into the low-order f-modes, and estimate the
signal-to-noise ratio for the future giant flare detection with advanced LIGO. By contrast, the torsional modes can be strongly
excited and may well be responsible for some of the observed QPO's in magnetar flares.

\section{The general formalism}
The giant flares release a significant fraction of the free magnetic energy stored
in their host magnetars. Two distinct mechanisms for this have been proposed:
1. Large-scale rearrangement of the internal field, facilitated by a major rupture of
the crust (Thompson \& Duncan 1995, 2001; we shall refer to it as the internal mechanism, IM), and 
2. Large-scale rearrangement of the magnetospheric field, facilitated 
by fast reconnection (Lyutikov 2006, Gill \& Heyl 2010; we shall refer to
it as the external mechanism, EM ). Both processes may well be at play:
the IM would likely serve as a trigger for the EM (however, as was argued in Lyutikov 2003, EM may also
be triggered by slow motion of the footpoints of a magnetospheric flux tube, leading
to a sudden loss of magnetostatic equilibrium). Observationally, the extremely
short, a few microseconds rise time of the 2004 giant flare in SGR 1806-20 (Hurley et al.~2004) gives
reason to believe that EM was at
play in that source: the IM operates on a much longer Alfven crossing timescale
of $~0.05$--$0.1$ seconds. The long timescale for the IM implies that it would not be efficient in
exciting the  f-modes
 which have frequencies of over a kHz; this was recently independently emphasized by
Kashiyama \& Ioka (2011). We shall come back to this point more quantitatively at the end of this section.   In what follows we will thus
concentrate on the mode excitation by the EM. The IM may well be efficient in exciting  lower-frequency
torsional  modes; however, as we discuss below, its coupling to these modes is of similar
strength to that of the EM, and  as we will show, both mechanisms may be
 efficient in exciting the large-amplitude
torsional modes.

During the large-scale EM event, the magnetic stresses at the stellar surface change instantaneously
(in several light-crossing timescales) by, at most\footnote{During the short timescale of the EM event no
new flux can emerge from the star.}, order 1. 
We shall characterize the change of the magnetic stress  by the 3 components
\begin{eqnarray}
\Delta T_{rr}&=&{B^2\over 4\pi} f_r(\theta, \phi),\nonumber\\
\Delta T_{r\theta}&=&{B^2\over 4\pi} f_{\theta}(\theta,\phi),\label{stress1}\\
\Delta T_{r\phi}&=&{B^2\over 4\pi} f_{\phi}(\theta, \phi),\nonumber
\end{eqnarray}
where $B$ is some characteristic value of the surface magnetic field and
$f_r$, $f_\theta$, and $f_\phi$ are functions are of order 1 in the strongest possible flares and
are smaller for the weaker flares. Consider now a normal
mode of the star with an eigenfrequency $\omega_n$ and a displacement wavefunction 
$\vec{\xi}_n(r, \theta, \phi)$. We treat the changing surface magnetic stress as an external 
perturbation acting on the mode. The Lagrangian of the free (pre-perturbation) mode is given by
\begin{equation}
L_{\rm free}(a_n,\dot{a}_n)={1\over 2} m_n \dot{a}_n^2-{1\over 2}m_n\omega_n^2 a_n^2,
\label{freeL}
\end{equation}
where 
$a_n$ is the generalized coordinate corresponding to the normal mode, $m_n$ is the 
effective mass given by
\begin{equation}
m_n=\int d^3 r \rho(\vec{r}) \vec{\xi}^2_n(\vec{r}),
\end{equation}
and $\rho(\vec{r})$ is the density.
The Lagrangian term characterising the mode's interaction with external stress
 is given by (cf.~section 2 of Levin 1998)
\begin{equation}
L_{\rm int}=a_n\int R^2 \vec{\xi}_n\cdot\vec{F}
\sin\theta  d\theta d\phi,
\end{equation}
where 
\begin{equation}
\vec{F}=\Delta T_{rr} \vec{e}_r+\Delta T_{r\theta}\vec{e}_{\theta}+\Delta T_{r\phi}
\vec{e}_{\phi},
\end{equation}
and the displacement $\vec{\xi}$ is evaluated at the radius of the star $R$.
The full Lagrangian for the $n$th mode is given by\footnote{We work in the linear regime, and don't
take into accounf the non-linear coupling between the modes. The mode amplitudes $\ll 1$ found at the 
end of our calculation indicate that this is a good approximation.}
\begin{equation}
L(a_n, \dot{a}_n)=L_{\rm free}+E_{\rm mag}\alpha_n {a_n\over R},
\label{fullL}
\end{equation}
where 
\begin{equation}
E_{\rm mag}={B^2R^3\over 4\pi} 
\end{equation}
is the characteristic energy stored in the star's magnetic field and
and $\alpha_n$ is the coupling coefficient given by
\begin{equation}
\alpha_n=\int \vec{\xi}_n(R, \theta, \phi)\cdot \vec{f}(\theta, \phi) \sin\theta d\theta d\phi,
\label{alphan}
\end{equation}
where
\begin{equation}
\vec{f}=f_r(\theta, \phi)\vec{e}_r+
        f_{\theta}(\theta, \phi)\vec{e}_{\theta}+f_{\phi}(\theta, \phi)\vec{e}_{\phi}
\label{alphan1}
\end{equation}
It is now trivial to find the motion resulting from the sudden introduction of the external stress at
moment $t=0$.
The coordinate $a_n$ oscillates as follows:
\begin{equation}
a_n(t)=\bar{a}_n\left[1-\cos\left(\omega_n t\right)\right],
\end{equation}
where the amplitude is given by
\begin{equation}
\bar{a}_n={\alpha_n E_{\rm mag}\over m_n\omega_n^2 R}.
\label{amplitude}
\end{equation}
The energy in the excited mode is given by
\begin{equation}
E_n={\alpha_n^2 E_{\rm mag}^2\over 2 m_n \omega_n^2 R^2}
\label{energy}
\end{equation}

We now briefly revisit the mode excitation by the IM. In this case, the interaction Lagrangian of a mode with
the magnetic field is described by the following volume integral:
\begin{equation}
L_{\rm int}=a_n\int d^3r  \vec{f}_L(\vec{r})\cdot\vec{\xi}_n(\vec{r}),
\label{volumeint}
\end{equation}
where $\vec{f}_L=\left[\nabla\times\vec{B}\right]\times\vec{B}$ is the lorentz force per unit volume.
Since $f_L\sim B^2/R$, one can see that the coupling of the internal field variation to the mode is
of the same order of magnitude as that of the external field variation, provided that the external and
internal fields are of the same order of magnitude. However, the IM mechanism acts on a much longer
 timescale\footnote{This timescale could be shorter if the superfluid neutrons are decoupled from the 
MHD (Easson \& Pethick 1979, van Hoven \& Levin 2008, Andersson, Glampedakis, \& Samuelsson 2009).
However, given the large fluid velocities expected in IM, this decoupling seems unlikely.}  $\tau_{\rm Alfven}\sim 0.1$s than the typical  f-mode period of $\tau_f\sim 0.0005$s, so the f-mode
oscillator would be adiabatically displaced without excitation of the periodic oscillations. 
One can show that the typical suppression factor of the IM relative to the EM excitation is {\it at least}
of order $2\pi \tau_{\rm Alfven}/\tau_f$ in the mode amplitude\footnote{This can be formalized by the following
 argument: consider a harmonic oscillator of proper frequency $\omega_0$, initially at rest, which is externally driven by
force $f(t)$. The amplitude of the induced oscillation at the proper frequency is proportional
to $\tilde{f}(\omega_0)$, the Fourier transform of $f(t)$ evaluated at $\omega_0$. For a step function, as expected
in EM, $\tilde{f}(\omega)\propto 1/\omega$. On the other hand, 
for a smooth pulse of duration $\tau$, as expected in IM, the Fourier transform is suppressed and scales
at most as $\tilde{f}(\omega)\propto (\omega \tau)^{-1}1/\omega$ when $\omega\tau
\gg 1$.
}. 
This factor is so large that even
if internal field was stronger than the external field by an order of magnitude, the IM excitation would
still be suppresed relative to the EM one. 

Is there a way around this suppression factor?  Potentially, IM could feature a collection of many localized MHD excitation, with the timescale for each one being determined by the Alfven-crossing time  of each of the excitation domain. If the domains were small enough, their
timescales could be more closely matched with the f-mode period (Melatos, private communications). 
However, in this case the magnitude of the
overlap integral in Eq.~(\ref{volumeint}) would be reduced by a factor $\sim (R/\Delta R)^3$, where 
$\Delta R$ is the characteristic size of the excited domain.  The domains would contribute incoherently to the amplitude of the excited mode, thus the contribution of an individual domain would have to be multiplied
by  $(R/\Delta R)^{3/2}$ in the (somewhat unlikely) 
limit where the active domains occupy the whole star. Thus, while the
timescale of the mini-flares could be well-matched with the f-mode period, their overall conrtribution
to the overlap integral in Eq.~(\ref{volumeint}) would be suppressed by $\sim (R/\Delta R)^{3/2}$. In the optimal case
that the mini-flares have the same timescale as the f-mode period, $R/\Delta R\sim \tau_{\rm Alfven}/\tau_f$. Therefore, the collection of mini-flares would not give us any gain in 
the mode excitation amplitude, as compared to the IM estimate given in the previous paragraph.

Two applications of the  formalism for the mode excitation by the  EM mechanism
 developed above are presented in the next two sections.

\section{f-modes and gravitational waves}
In order to estimate an effective f-mode mass,
we have computed the $l=2$ f-mode displacement functions for a neutron star\footnote{We constructed our neutron star model using the equation of state from Douchin \& Haensel (2001) and Haensel \& Pichon (1994). In calculating the f-mode we treated the whole star as a fluid, neglecting the effects of bulk- and shear moduli.} in the Cowling approximation.
Convenient scalings are
\begin{eqnarray}
m_n&=&q_M M,\nonumber\\
\omega_n^2&=&q_{\omega} {GM\over R^3},\nonumber\\
\xi_r(R, \theta, \phi)&=&a_n Y_{2m}(\theta, \phi).\nonumber\\
\end{eqnarray} 
In our fidutial model $q_M=0.046$, where we have normalised the mode wavefunction so that
$\vec{e}_r\cdot \vec{\xi}_{2m}(R,\theta,\phi)=Y_{2m}(\theta,\phi)$. Our reference number $q_{\omega}=1.35$ was obtained using a fitting formula for fully relativistic f-mode frequencies\footnote{We are not being consistent in, on the one hand, using the Cowling approximation for a Newtonian star
 to determine the effective mode mass, but  on the other hand using the published 
relativistic calculations for the mode frequencies. Normally, Newtonian calculations would be 
sufficient, given the many unknown
details of the flare and the many poorly constrained parameters we'd already introduced into the model, and
the formalism we developed in the previous section is manifestly Newtonian (but can be generalized to relativistic regime if the need arises). However, as we show below, the
signal-to-noise ratio for the gravitational-wave detection is very sensitive to the mode frequency, and 
therefore we try to be accurate in characterizing these frequencies.}  
from Andersson \& Kokkotas (1996). The amplitude of the f-mode is given by
\begin{equation}
{\bar{a}_{2m}\over R}={\alpha_{2m}\over q_m q_{\omega}}{E_{\rm mag}\over E_{\rm grav}},
\label{ampfmode}
\end{equation}
 where 
\begin{equation}
E_{\rm grav}=\frac{GM^2}{R}
\end{equation}
is of the same order as the gravitational binding energy of the neutron star.
For our fidutial model with $B=10^{15}$G, $M=1.4 M_{\odot}$, and $R=10$km, 
we get ${\bar{a}_{2m}/ R}\sim 3\times 10^{-6}\alpha_{2m}$.
The energy in the f-mode is
\begin{equation}
E_f = \frac{\alpha_{2m}^2}{2 q_m q_{\omega}}\frac{E_{\rm mag}^2}{E_{\rm grav}}\sim 1.5\times 10^{-6}
\alpha_{2m}^2 E_{\rm mag}
\end{equation}
for our fidutial parameters.
This energy is drained from the star primarily through emission of gravitational waves. The total amount of energy carried by gravitational waves is therefore 
\begin{equation}
E_{\rm GW} = E_f = \frac{2\pi^2 d^2 f^2 c^3}{G}\int_{-\infty}^{\infty} \langle h^2\rangle   dt
\label{GW_energy}
\end{equation}
where $f = \omega_n/2\pi$ is the f-mode frequency in Hz,  $\langle h^2\rangle$ is the
direction and polarisation averaged value of the square of the gravitational-wave strain $h$ as measured by observers at distance $d$ from the source. This expression allows us to estimate the expected signal-to-noise ratio for ground based gravitational wave interferometers (cf.~Abadie et al 2010). One can use the fact
that nearly all the gravitational-wave signal is expected to arrive in a narrow-band around the f-mode
frequency, and that the signal form (the exponentially-decaying sinusoid) is known. 
The Wiener-filter expression for
the signal-to-noise can be written as
\begin{eqnarray}
 \frac{S}{N} &\approx& \left[  \frac{1}{S_h (f)} \int_{-\infty}^{\infty} \vert \tilde{h}^2 (f') \vert df' \right]^{1/2}\\
 &\sim& \left[ \frac{G}{2\pi^2 c^3} \frac{E_f}{S_h(f) f^2 d^2} \right]^{1/2}
\end{eqnarray}
where $\tilde{h} (f)$ is the Fourier transform of the time-dependent gravitational wave strain $h (t)$. As is
standard for narrow-band signal, we have used Parseval's theorem to convert the integral over $f$ to the integral over $t$ from Eq. (\ref{GW_energy}),
and, following Abadie et al.~2010, we have approiximated $h^2$ with the average $\langle h^2\rangle$. 
At frequencies of a few kHz the spectral density, $S_h(f)$, of the ground based detectors like Advanced LIGO and Virgo is dominated by shot-noise and is proportional to $f^2$. This makes the signal-to-noise ratio for observations of magnetar f-modes excited in giant flares particularly sensitive to frequency ($\propto f^{-3}$). For Advanced LIGO we find
\begin{eqnarray}
 \frac{S}{N} \approx 0.07 ~\alpha_{2m}\left( \frac{2000 ~\rm{Hz}}{f} \right)^3 \left( \frac{B}{10^{15} ~\rm{G}} \right)^2 \left( \frac{1 ~\rm{kpc}}{d} \right) \times \nonumber \\ 
 \left( \frac{R}{10 ~\rm{km}} \right)^2 \left( \frac{0.07 ~M_{\odot}}{m_n} \right)^{1/2}
\label{signaltonoise}
\end{eqnarray}
Here we used tabulated\footnote{These sensitivity curves represent the incoherent sum of principal sources of noise as they are currently understood.} $S_h(f)$ from the LIGO document LIGO-T0900288, which gives \newline $S_h(f) = 8.4 \cdotp 10^{-47}
\hbox{Hz}^{-1} \left( f/2000 ~\rm{Hz}\right)^2$ for the shot-noise dominated part of the curve.

\section{Torsional modes}
Intuitively, one expects torsional modes to be strongly excited during the magnetar flares (Duncan 1998),
since it is the free energy of the twisted magnetic field that is being released.
These have much lower proper frequencies than the f-modes (with the fundamental believed to be in the
range $10--40$Hz, see Steiner 
\& Watts 2009 and references therein), which can be well-matched to the Alfven
frequencies inside the star. Thus both EM and IM are likely to play a role in the torsional mode excitation.
Here, we consider the EM explicitly but keep in mind that IM would give a similar answer.

For the torsional modes in the crust, the displacement is given by
\begin{equation}
\vec{\xi}_{nlm}(r,\theta,\phi)=g_n(r)\vec{r}\times\nabla Y_{lm}(\theta,\phi),
\label{tordisp}
\end{equation}
and it is convenient to normalise the wavefunctions so that $g_n(R)=1$. Here $n=0,1,...$ is the number
of radial nodes.

With this normaliation, the effective mode mass $m_{nlm}\sim m_{\rm crust}\sim 0.01M$, and from 
Eq.~(\ref{amplitude}) one gets for the mode amplitude normalized by the star radius:
\begin{equation}
{a_{nlm}\over R}\sim 0.01\alpha_{nlm} \left({B\over 10^{15}\hbox{G}}\right)^2{R\over 10\hbox{km}}
{0.014M_{\odot}\over m_{nlm}}\left({100\hbox{Hz}\over f}\right)^2.
\label{anlm}
\end{equation}
Thus we see that for a reasonable range of parameters it is feasible that the crustal torsional modes would
be strongly excited by a giant flare.

\subsection{Magnetic modes}
Recently, Kashiyama \& Ioka (KI, 2011) suggested that  certain types of MHD modes that may be strongly excited during a giant flare, are coupled to gravitational radiation and may therefore 
become an interesting source for advanced LIGO. KI focus on the polar modes of Sotani \& Kokkotas (2009); the MHD modes found by
Lander \& Jones (2011a,b) also satisfy some of the KI's criteria.

While interesting, this idea has potential caveats that need further investigation. 
The polar, axially-symmetric modes of Sotani \& Kokkotas (2009) exist in idealized
polytropic stars; in a real neutron star the frequency of such modes would likely be
many times greater due to the strong stable stratification (Goldreich \& Reisenegger 1992, Reisenegger \& Goldreich
1992) and their nature would  likely be closer to that of the g-modes. More importantly, KI assume that the
oscillations are long-lived, $\sim10^7$ oscillation periods. However, MHD modes are notoriously capricious. While the idealized modes of Sotani \& Kokkotas (2009)
and Lander \& Jones (2011a,b) are protected by symmetry, the global magnetic modes in more realistic 
configurations may couple to a variety of localized Alfven-type modes, and may thus be quickly
damped via phase mixing and resonant absorption (Goedbloed \& Poedts 2004, van Hoven \& Levin 2011). Thus, in our view, 
there is currently no compelling reason to believe that
the magnetic modes can be substantially longer lived than the observed magnetar QPOs.

\section{Discussion}
In this paper, we have computed the excitation of the f-modes and crustal tosional neutron-star modes
by a giant flare. Corsi \& Owen (2011) recently computed the magnetic energy that can be released during the
flare\footnote{These analytical calculations necessarily make simplifying assumptions about the structure of
an equilibrium magnetic field inside the magnetar, but they are likely to give correct order-of-magnitude values.}, and found values comparable to $E_{\rm mag}$. However, in this work we showed that
 only a small fraction of the released flare energy is converted into the f-modes, and that the
associated gravitational-wave emission is correspondingly weaker than has been previously hoped
(cf.~Abadie et al.~2010 and Corsi \& Owen 2011).
From Eq.~(\ref{signaltonoise}), our fiducial model does not look promising for future advanced LIGO detection of a giant flare, 
even if $\alpha_{2m}\sim 1$,
i.e.~if the released electro-magnetic energy is of order of the total magnetic energy of the star, $\sim 10^{47}$erg (the most energetic of the 3 observed giant flares released few$\times 10^{46}$erg). 
However, if the surface field
is significantly larger than $10^{15}$G and/or the star is greater than $10$km in radius (which would reduce
the f-mode frequency and increase the contact surface area), then one can become  more hopeful
about the potential detection.

On the other hand, we have seen that there is no difficulty in exciting the crustal torsional modes
to a large amplitude. Whether or not this leads to the observed quasi-periodic oscillations
in the flare's tail (Israel et al.~2005, Strohmayer \& Watts 2005, Watts \& Strohmayer 2006)
 depends crucially on the dynamics of
hydromagnetic coupling between the crustal modes and  the Alfven modes of the magnetar core
(Levin 2006, 2007, van Hoven \& Levin 2011, Gabler et al.~2011, Colaiuda \& Kokkotas 2011).

\section{Acknowledgements}
We thank Ben Owen for useful discussions, and Andrew Melatos and Peter Kamus for helpful comments on the initial draft of this paper submitted for  an internal LIGO Scientific Collaboration review. 
This research was suppoted, in part, by Leiden Observatory and Lorentz Institute through internal grants. MvH thanks Monash School of Physics, where this research was completed, for hospitality 
during his extensive visit.

\section*{References}
\begin{footnotesize} \noindent
Abadie, J., et al.~(LIGO Scientific Collaboration, VIRGO), 2010, arXiv1011.4079\\ 
Abbott, B.~P., et al., 2008, ApJ, 681, 1419\\
Abbott, B.~P., et al., 2009a, Reports on Progress in Physics, 72, 076901\\
Abbott, B.~P., et al., 2009b, ApJ Letters, 701, L68\\
Acernese, F., et al., 2008, Classical and Quantum Gravity, 25, 114045\\
Andersson, N., Glampedakis, K., \& Samuelsson, L., 2009, MNRAS, 396, 894\\
Andersson, N., \& Kokkotas, K.~D., 1996, Phys.~Rev.~Letters, 77, 4134\\
Barat C. et al., 1983, A\& A, 126, 400\\
Colaiuda, A., \& Kokkotas, K.~D., 2011, arXiv1012.3103\\
Corsi, A., \& Owen, B., 2011, arXiv1102.3421\\
Douchin, F., \& Haensel, P., 2001, A\&A, 380, 151\\
Duncan, R.~C., 1998, ApJ Letters, 498, 45\\
Gill, R., \& Heyl, J.~S., 2010, MNRAS, 407, 1926\\
Gabler, M., Cerda-Duran, P., Font, J., Muller, E., \& Stergioulas, N., 2011, MNRAS Letters, 410, 37\\
Gill, R., \& Heyl, J.~S., 2010, MNRAS, 407, 1926\\
Goedbloed, J.~P., \& Poedts, S., 2004, Principles of Magnetohydrodynamics (Cambridge University Press)\\
Goldreich, P., \& Reisenegger, A., 1992, ApJ, 395, 250\\
Easson, I., \& Pethick, C.~J., 1979, ApJ, 227, 995\\
El-Mezeini, A.~M., \& Ibrahim, A.~I., 2010, ApJL, 721, 121\\
Haensel P., \& Pichon, B., 1994, A\&A, 283, 313\\
http://www.ioffe.ru/astro/NSG/NSEOS/\\
Haensel P., Potekhin A.~Y., Yakovlev D.~G., 2007. Neutron Stars 1: Equation of State
and Structure (New York: Springer)\\
Hurley K., Boggs S. E., Smith D. M., Duncan R. C., Lin R.,
Zoglauer A., Krucker S., Hurford G., Hudson H., Wigger C., Hajdas W., Thompson C., Mitrofanov I., Sanin A., Boynton W., Fellows C., von Kienlin A., Lichti G., Rau A., 2005, Nature, 434, 1098\\
Hurley K., Cline T., Mazets E., Barthelmy S., Butterworth P., Marshall F., Palmer D., Aptekar R., Golenetskii S., Il’Inskii V., Frederiks D., McTiernan J., Gold R., Trombka J., 1999, Nature, 397, 41\\
Israel G.~L. et al., 2005, ApJ, 628, L53\\
Kalmus, P., Cannon, K.~C., Marka, S., \& Owen, B., 2009, Phys.~Rev.~D, 80, 042001\\
Kashiyama, K., \& Ioka, K., 2011, submitted to Phys.~Rev.~D,  arXiv:1102.4830\\
Lander, S.~K., \& Jones, D.~I., 2011a, MNRAS in press, arXiv1009.2453\\
Lander, S.~K., \& Jones, D.~I., 2011b, MNRAS in press, arXiv1010.0614\\ 
Levin, Y., 1998, Phys.~Rev.~D., 57, 659\\
Levin, Y., 2006, MNRAS Letters, 410, 37\\
Levin, Y., 2007, MNRAS, 377, 159\\
Lyutikov, M., 2003, MNRAS, 346, 540\\
Lyutikov, M., 2006, MNRAS, 367, 1594\\
Mazets E. P., Golentskii S. V., Ilinskii V. N., Aptekar R. L.,
Guryan I. A., 1979, Nature, 282, 587\\
Mereghetti S., 2008, Astron. Astrophys. Rev., 15, 225\\
Reisenegger, A., \& Goldreich, P., 1992, ApJ, 395, 250\\
Sotani, H., \& Kokkotas, K.~D., 2009, MNRAS, 395, 1163\\
Steiner W., Watts A.~L., 2009, Phys.~Rev.~Letters, 103r1101S\\
Strohmayer, T.~E. \& Watts, A.~L., 2005, ApJ, 632, L111\\
Thompson, C., \& Duncan, R.~C., 1995, MNRAS, 275, 255\\
Thompson, C., \& Duncan, R.~C., 2001, ApJ, 561, 980\\
van Hoven, M., \& Levin, Y.~2008, MNRAS, 391, 283\\
van Hoven, M., \& Levin, Y., 2011, MNRAS, 410, 1036\\
Watts, A.~L. \& Strohmayer T.~E., 2006, ApJ, 637, L117\\
Woods P. M., Thompson C., 2006, Soft gamma repeaters and anomalous X-ray pulsars: magnetar candidates. pp 547–586\\
\end{footnotesize}




\end{document}